\documentclass{PoS}

\def\MeV {\mathop{\hbox{MeV}}}

\def\Im {\mathop{\hbox{Im}}}

\usepackage{amsmath} 
\usepackage{dsfont}  
\usepackage{bm}      

\def\beq{\begin{equation}}
\def\eeq{\end{equation}}
\def\beqs#1\eeqs{\beq\begin{split} #1 \end{split}\eeq}

\def\comment#1{}

\def\av#1{ \left\langle #1 \right\rangle }

\graphicspath{{figures/}}

\title{QCD at imaginary chemical potential with Wilson fermions}

\ShortTitle{QCD at imaginary chemical potential with Wilson fermions}

\author{\speaker{Andrei Alexandru}\\
        Physics Department, The George Washington University, Washington, DC, 20052, USA\\
        E-mail: \email{aalexan@gwu.edu}}

\author{Anyi Li\\
        Institute for Nuclear Theory, University of Washington, Seattle, 98195, USA\\
        E-mail: \email{anyili@uw.edu}}

\abstract{
We investigate the phase diagram in the temperature, imaginary chemical
potential plane for QCD with three degenerate quark flavors using
Wilson type fermions. While more expensive than the staggered fermions 
used in past studies in this area,
Wilson fermions can be used safely to simulate systems with three
quark flavors. In this talk, we focus on the (pseudo)critical line 
that extends from $\mu=0$ in the imaginary chemical potential plane, 
trace it to the Roberge-Weiss line, and determine its location relative to the 
Roberge-Weiss transition point. In order to
smoothly follow the (pseudo)critical line in this plane we perform 
a multi-histogram reweighting in both temperature and chemical potential. 
To perform reweighting in the chemical potential we use the 
compression formula to compute the determinants exactly. Our results 
are compatible with the standard scenario.
}

\FullConference{31st International Symposium on Lattice Field Theory - LATTICE 2013\\
		July 29 - August 3, 2013\\
		Mainz, Germany}

\begin{document}

\section{Introduction}

The phase diagram of QCD at non-zero baryon density is interesting for both
experimental and theoretical reasons. In the temperature range around the
deconfining transition the non-perturbative effects are important and 
lattice QCD could provide important input. However, direct simulations
are not yet possible due to the notorious {\em sign problem}. On the other
hand, the phase diagram of QCD at imaginary chemical potential can be 
determined using lattice QCD since the integration measure becomes real.
It is then possible to map out the phase diagram for $\mu^2<0$ and then
use analyticity or reweighting to inform us about the phase diagram for $\mu^2>0$.

For imaginary chemical potential, the integration measure is real due to the 
$\gamma_5$-hermiticity of the fermionic matrix, i.e.
\beq
M(U,\mu)^\dagger = \gamma_5 M(U,-\mu^*)\gamma_5 \,.
\eeq
The other important property constraining the phase diagram in the imaginary
chemical potential plane is the behavior of the integrand in the grand canonical 
partition function,
\beq
Z_\text{GC}(T,V,\mu) = \int {\cal D}U e^{-S_g(U)} \det M(U,\mu) \,,
\eeq
under the $Z(3)$-transformation $U\to U_\pm$ with
\beq
[U_\mu(\bm x, t)]_\pm = \begin{cases} 
U_\mu(\bm x, t)e^{\pm i\frac{2\pi}3} & \text{if $t=N_t-1$ and $\mu=4$},\\
U_\mu(\bm x, t) & \text{otherwise.}
\end{cases}
\eeq
The gauge action $S_g$ and the Haar measure ${\cal D}U$ are invariant under
this transformation and the effect on the fermionic part can be viewed as a
shift in the chemical potential, 
\beq
\det M(U_\pm, \mu) = \det M(U, \mu\pm i\frac{2\pi}3 T)\,.
\label{eq:1.4}
\eeq
This leads to the periodicity of the grand canonical partition function
\beq
Z_{GC}(T,V,\mu)= Z_{GC}(T,V,\mu \pm i\frac{2\pi}3 T)\,.
\eeq

Charge conjugation symmetry relates the probability distribution of two
gauge configurations that are complex conjugated. The gauge action and the
integration measure are symmetric under the transformation, whereas the fermionic matrix
satisfies
\beq
\det M(U^*, \mu) = \det M(U, \mu^*)^* \,.
\eeq
This implies a $Z(2)$ symmetry when the chemical potential is $\mu/T\in\{i\pi, -i\pi/3, i\pi/3\}$.
For $\mu/T=i\pi$ the configurations $U$ and $U^*$ have equal probability, whereas for 
$\mu/T=\pm i\pi/3$ we have
\beq
P_{\pm i\pi/3}(U) = P_{\pm i\pi/3}((U^*)_\mp)\,,
\eeq 
where
\beq
P_{\mu/T}(U) \equiv \frac1{Z_\text{GC}(T,V,\mu)} e^{-S_g(U)} \det M(U,\mu)
\eeq
is the probability density for configuration $U$. This $Z(2)$ symmetry was studied
by Roberge and Weiss~\cite{Roberge:1986mm}. They found this symmetry is spontaneously
broken at high temperatures, whereas at low temperatures it is restored. 

\begin{figure}
\begin{center}
\includegraphics[height=1.8in]{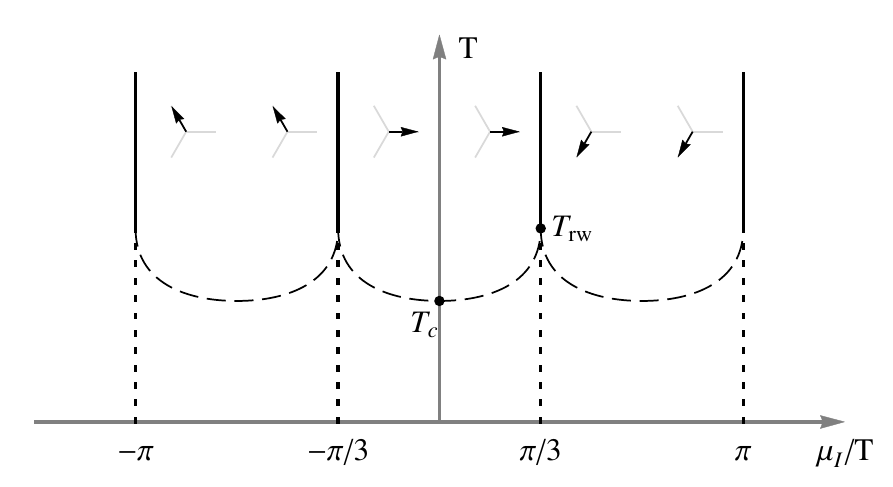} 
\raisebox{-0.13cm}{\includegraphics[height=1.8in]{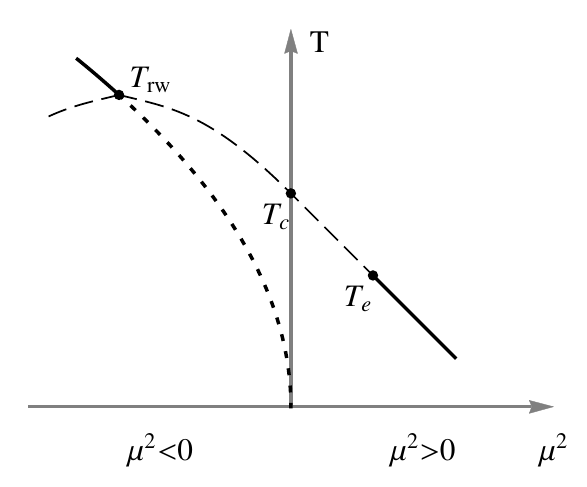}} 
\end{center}
\caption{The phase diagram in the imaginary chemical potential plane. Solid lines indicate
the first order phase transition and dotted lines cross-overs. The thin dashed lines represent
either rapid crossovers or real transitions depending on the number of quark flavors and their
mass.}
\label{fig:1}
\end{figure}

To better understand this behavior recall that at high temperature the Polyakov loop is {\em ordered}
and for $\mu=0$ the determinant favors the configurations with $\arg P\approx 0$.
Using Eq.~\ref{eq:1.4} this implies that for $\mu/T=\pm i2\pi/3$ the Polyakov loop
will be aligned differently, i.e. $\arg P\approx \mp2\pi/3$. The boundaries of these
regions are the Roberge-Weiss lines $\mu/T\in\{i\pi, -i\pi/3, i\pi/3\}$. When we cross
these lines at high temperatures the alignment of the Polyakov lines changes abruptly
and we have a first order phase transition. At low temperatures the transition is smooth
and we have a cross-over. The standard scenario is depicted in Fig.~\ref{fig:1}. 
The temperature where the Roberge-Weiss line changes from  first order to cross-over 
is denoted with $T_\text{rw}$. The expectation is that the Roberge-Weiss transition
point is connected with the (pseudo)critical transition point at zero density, $T_c$, which 
in turn is connected with the critical point at real imaginary chemical potential, $T_e$.
Whether the lines connecting these points are phase transitions or cross-overs depends
on the number of quarks in the system and their mass. 

\begin{figure}
\begin{center}
\includegraphics[height=1.8in]{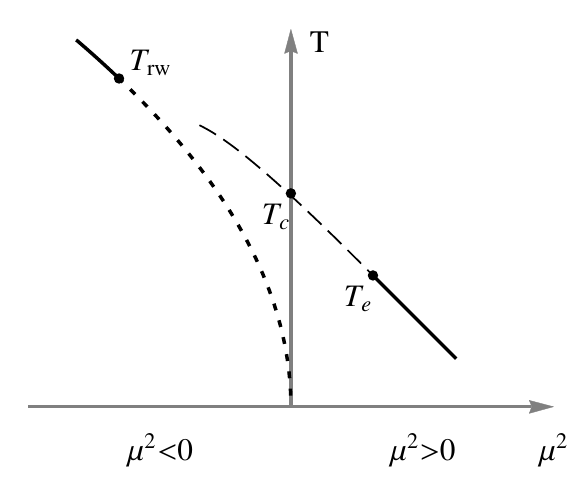} 
\raisebox{-0.0cm}{\includegraphics[height=1.8in]{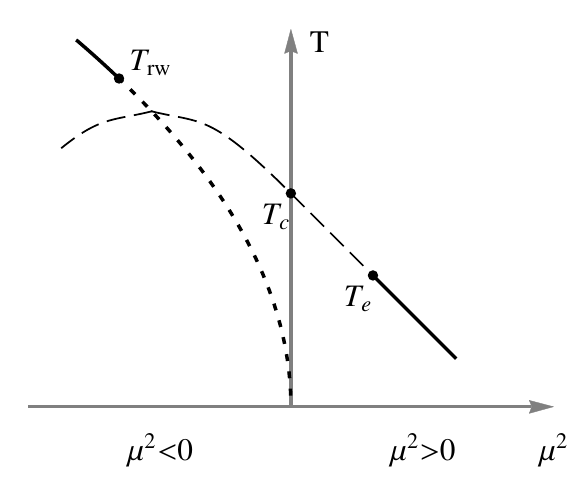}} 
\end{center}
\caption{Alternative scenarios: the (pseudo)critical line extending from $\mu=0$ could terminate before
touching the Roberge-Weiss line (left) or intersect it at a temperature different than $T_\text{rw}$.}
\label{fig:2}
\end{figure}

This is not the only logically possible scenario; alternative possibilities are depicted in Fig.~\ref{fig:2}:
the (pseudo)critical line extending from zero density might terminate before intersecting the
Roberge-Weiss line or intersect it at a temperature different than $T_\text{rw}$. 
It is then important to map out the phase diagram at imaginary chemical potential using 
direct simulations. The phase diagram at imaginary chemical potential was investigated for QCD with
two degenerate quark flavors~\cite{deForcrand:2002ci,DElia:2009qz,DElia:2009tm}, 
three flavors~\cite{deForcrand:2010he}, and 
four flavors~\cite{DElia:2002gd,DElia:2004at,DElia:2007ke}
using staggered fermions. For $N_f=2,3$ staggered simulations use the standard determinant root
technique and cross-checks with simulations using Wilson type fermions are required to remove
any possible doubts. For $N_f=2$ this was done by Nagata and Nakamura~\cite{Nagata:2011yf}.
In this talk we focus on the $N_f=3$ case.

\section{Technical details}

In this study we use gauge configurations generated using the Iwasaki gauge action and clover fermions.
To determine the structure of the phase diagram we focus on the region $1.65\leq \beta \leq 1.73$
and $\mu_I/T \in [0,\pi/3]$. To reduce the number of ensembles needed in our investigation we
used multi-histogram reweighting~\cite{Ferrenberg:1989ui}. Since we plan to trace out the 
(pseudo)critical line extending from zero density into the imaginary chemical potential plane,
we need to do a reweighting both in $\beta$ (temperature) and $\mu$. Schematically this 
can be achieved by introducing a reweighting factor $\alpha(U)$ via
\beq
\left\langle O(U) \right\rangle_{\beta,\mu} = \frac{\left\langle O(U) \alpha(U)\right\rangle_{\beta_0,\mu_0}}{\left\langle \alpha(U)\right\rangle_{\beta_0,\mu_0}}\,,
\eeq
where
\beq
\alpha(U)\equiv e^{-(\beta-\beta_0) S_g(U)}\frac{\det M(U,\mu)}{\det M(U,\mu_0)} \,.
\eeq
To compute the reweighting factor we need to compute a ratio of determinants. Our approach is to
compute this ratio exactly using determinant compression method~\cite{Alexandru:2010yb,Nagata:2010xi}.
The advantage of the compression method is that once the compressed matrix is diagonalized, we can 
compute the determinant for arbitrary values of the chemical potential as long as the other parameters
are kept fixed. Thus,
to facilitate this calculation we need to fix the value of the bare mass parameter $\kappa$ 
and the clover term $c_\text{sw}$. One of the goals for our study is to compare the reweighting
results to results derived in canonical simulations~\cite{Alexandru:2005ix}. We use the values that
correspond to $T=0.87 T_c$ in our $N_f=3$ study~\cite{Li:2011ee}, $\kappa=0.13825$ and $c_\text{sw}=1.89374$.
This corresponds to a pion mass of $m_\pi\approx 750\MeV$. 

\begin{figure}
\begin{center}
\includegraphics[height=2.2in]{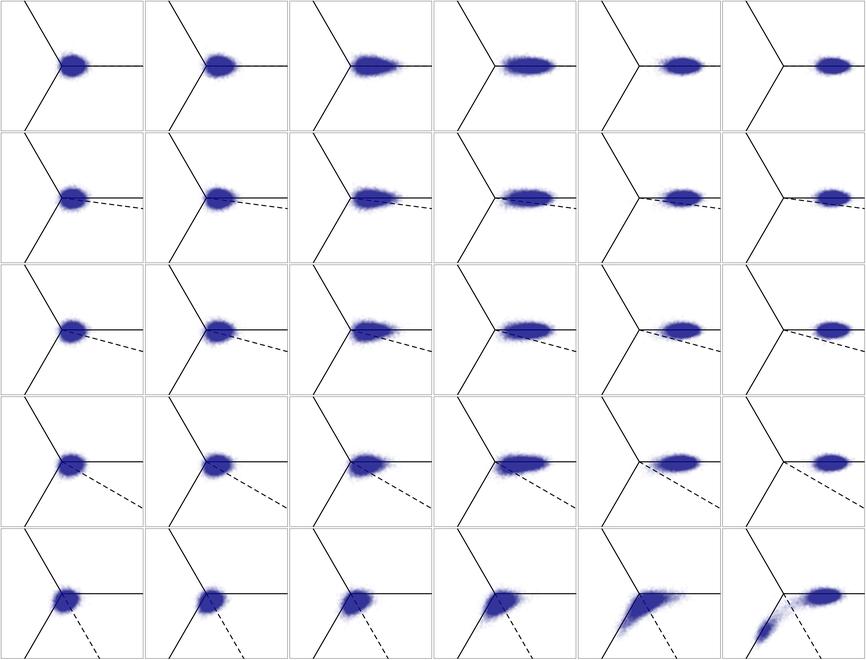} 
\hskip0.3in\raisebox{-0.3in}{\includegraphics[height=2.6in]{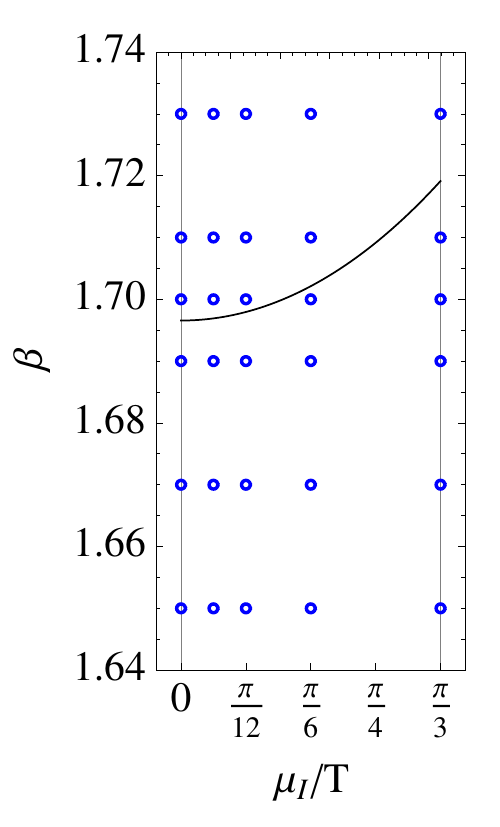}} 
\end{center}
\caption{Left: Polyakov loop distribution for $\mu_I/T=0$, $\pi/24$, $\pi/12$, $\pi/6$, and $\pi/3$ (top to bottom) and
$\beta=1.65$, $1.67$, $1.69$, $1.70$, $1.71$, and $1.73$ (left to right). The dashed line in each plot indicates the 
direction of $e^{-i\mu_I/T}$ in the complex plane. Right: the parameters of the generated ensembles
and the (pseudo)critical curve as determined from this study.}
\label{fig:3}
\end{figure}

In Fig.~\ref{fig:3} we show the distribution of the Polyakov loops and the positions in the 
phase diagrams for the ensembles used in this study. Each ensemble has about $20,000$ configurations
of size $6^3\times 4$. Note that the Polyakov loop prefers the
real sector except for the ensembles generated at $\mu_I/T=\pi/3$. This is the Roberge-Weiss line
and the $Z(2)$ symmetry is apparent. Note also that for $\beta=1.73$ the distribution of the Polyakov
loops is bimodal indicating that the symmetry is spontaneously broken, whereas for lower temperatures 
the Polykov loops follow a unimodal distribution signaling a restoration of the symmetry. 

\section{Numerical results}

\begin{figure}[!t]
\hbox{\kern0.5in 
\includegraphics[height=2.0in]{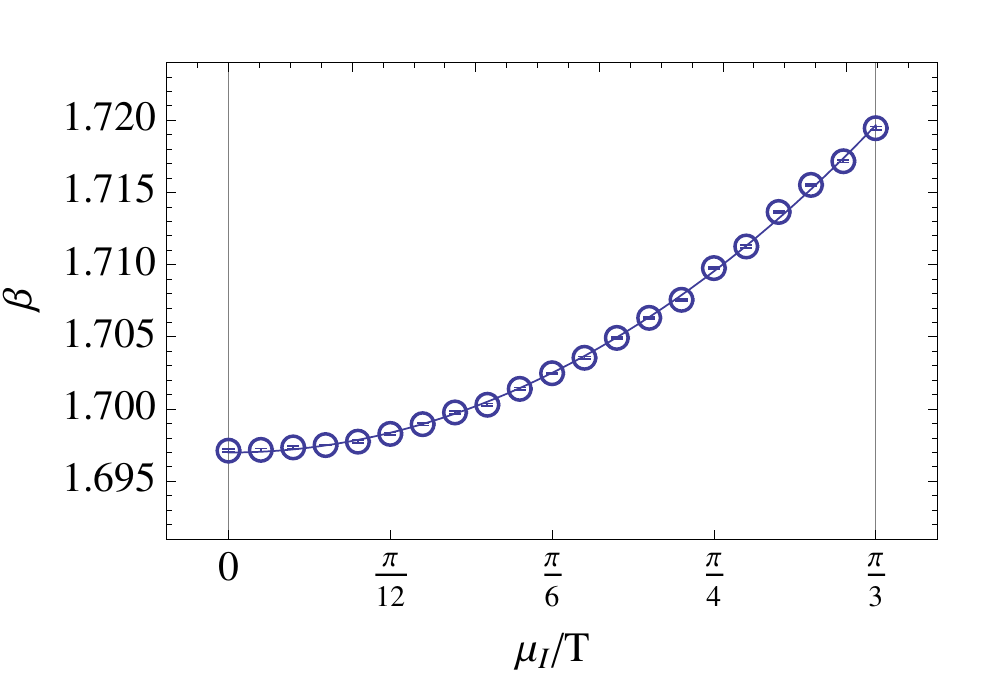} 
\kern -0in\raisebox{-0.05in}{\vbox{
\includegraphics[height=1.0in]{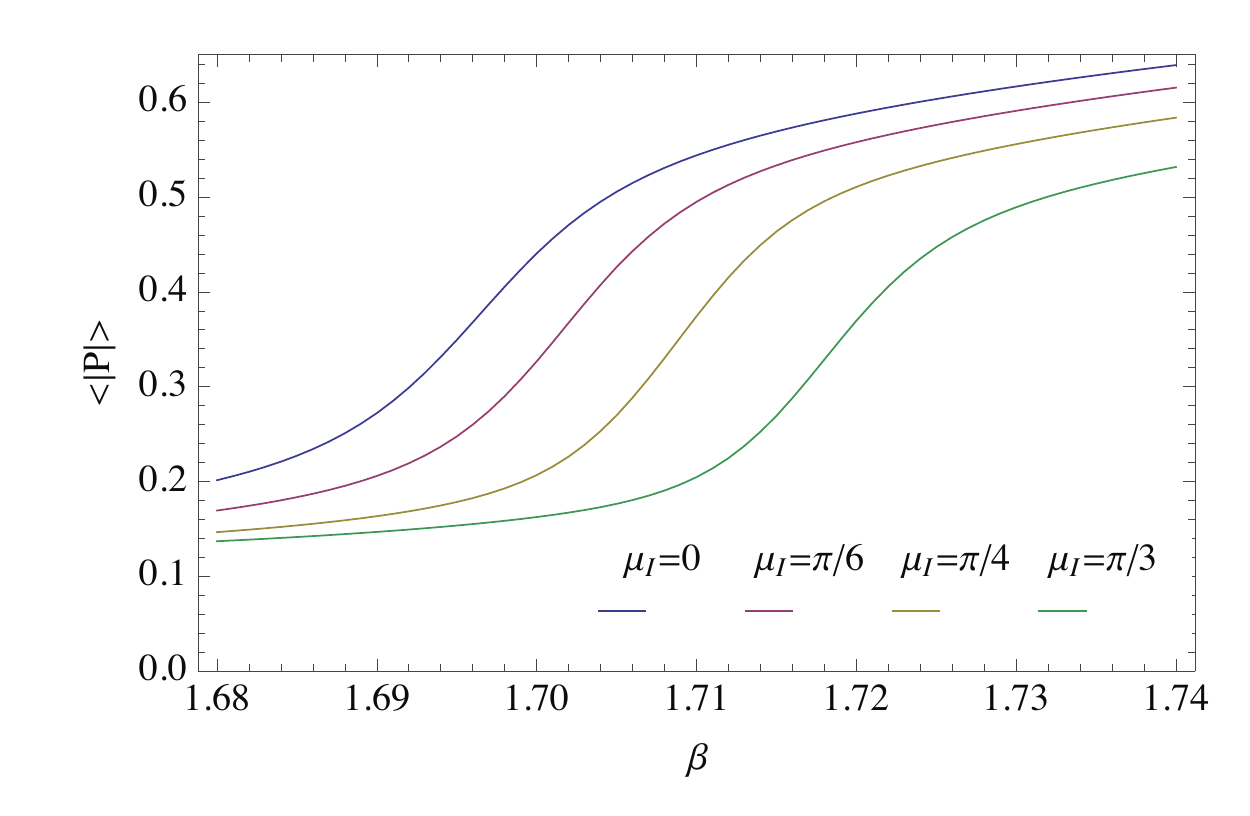}\\ 
\includegraphics[height=1.0in]{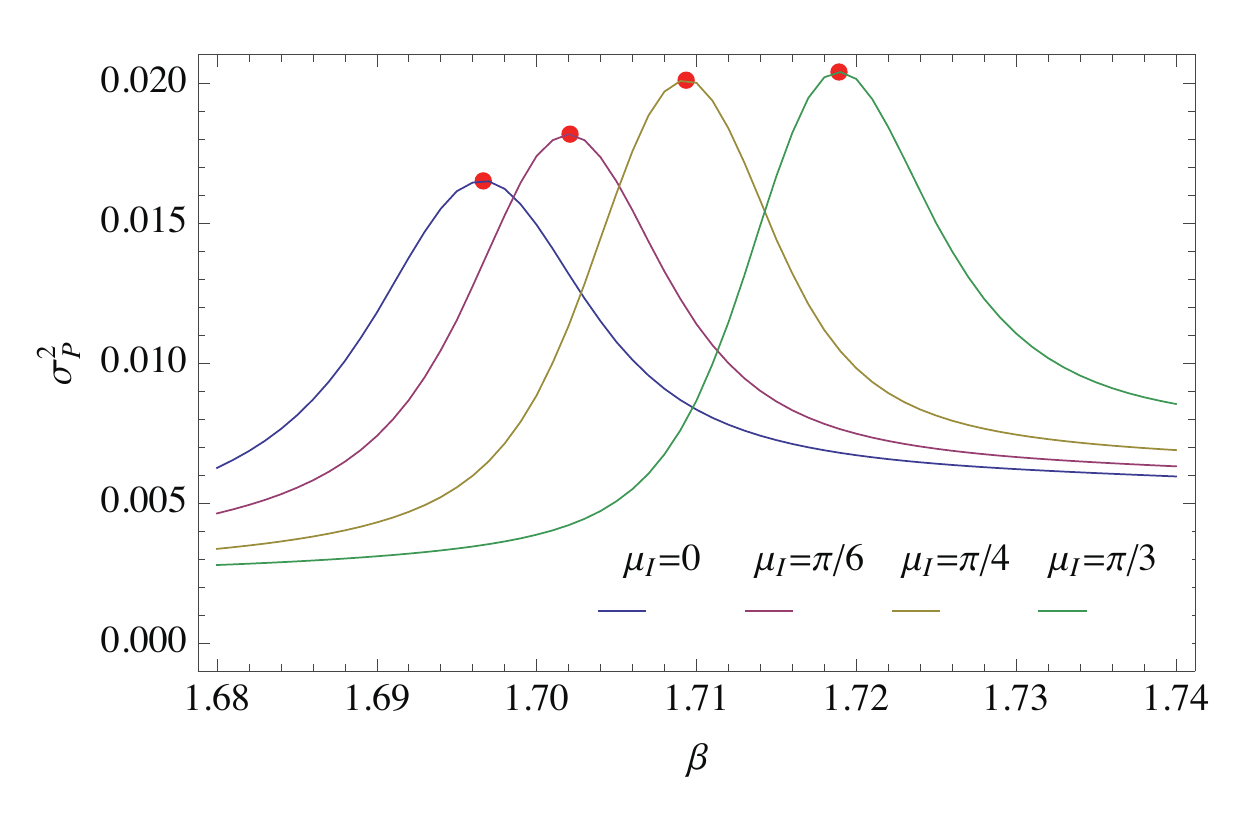}}}
}
\caption{Left: the (pseudo)critical temperature as a function of imaginary chemical potential. 
Right: Absolute value of the Polyakov loop as a function of temperature for fixed $\mu_I/T$ (top)
and its susceptibility (bottom). }
\label{fig:4}
\end{figure}

To map out the phase diagram we determine the (pseudo)critical line extending from $\mu=0$ and the
transition point on the Roberge-Weiss line. The first task can be accomplished by monitoring the
Polyakov line magnitude as we increase the temperature for fixed $\mu_I/T$. In the right panel
of Fig.~\ref{fig:4} we show that Polyakov loop curves for $\mu_I/T=0$, $\pi/6$, $\pi/4$, and $\pi/3$.
Note that the transition moves to higher temperature as we increase the imaginary chemical potential.
To better pinpoint the transition point we determine the Polyakov loop susceptibility and locate the
transition temperature at the point where the susceptibility reaches its maximal value. This is 
indicated in the lower figure from the right panel of Fig.~\ref{fig:4}. Using reweighting we can
trace the transition temperature as a function of $\mu_I$. The results are presented in the 
left panel of Fig.~\ref{fig:4}. The curve is well described by the following function:
\beq
\beta_c \left(\frac{\mu_I}T \right) = c_0 + c_2 \left(\frac{\mu_I}T \right)^2 + c_4 \left(\frac{\mu_I}T \right)^4 \,,
\eeq
with $c_0=1.69658(3)$, $c_2=0.02014(6)$, and $c_4=0.00040(6)$.

\begin{figure}[!t]
\begin{center}
\includegraphics[height=1.9in]{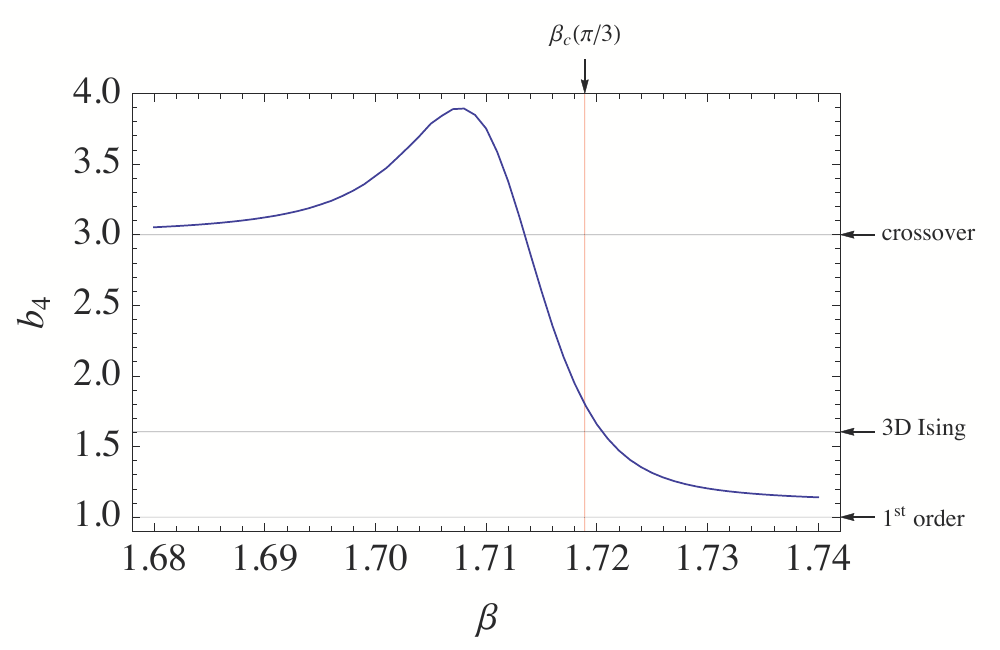} 
\includegraphics[height=1.9in]{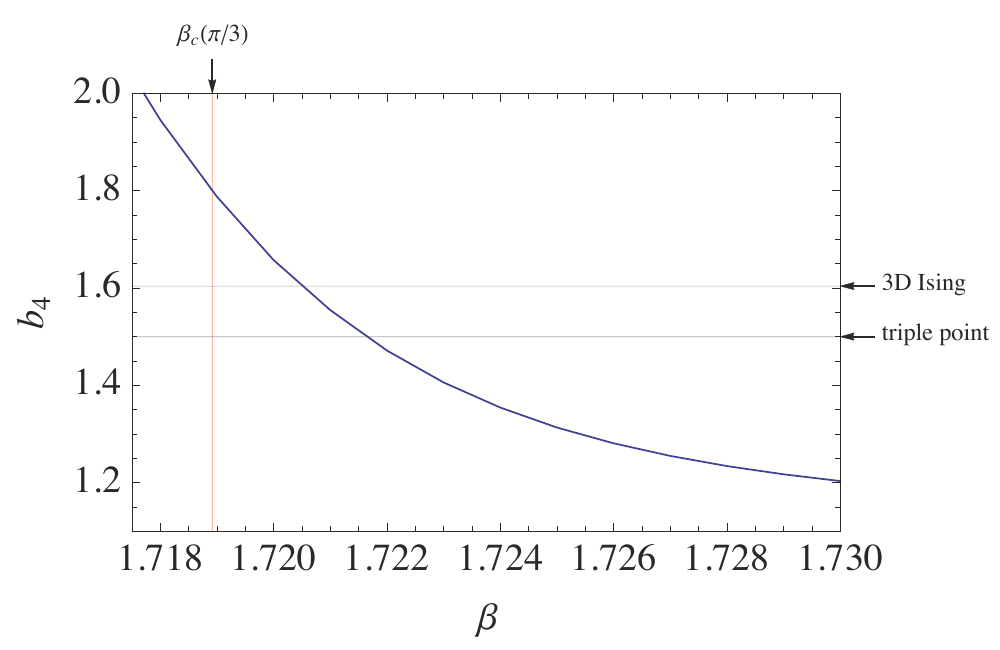} 
\includegraphics[width=6in]{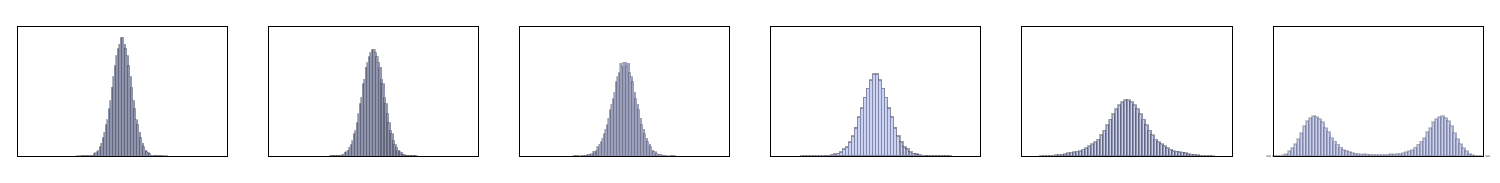}
\end{center}
\caption{Top: the Binder cumulant for the phase of the Polyakov loop along the line $\mu_I/T=\pi/3$ (left)
and a detailed look at the plot in the vicinity of $T_\text{rw}$ (right). 
We indicated on the graph the intersection between the (pseudo)critical line and
Roberge-Weiss line, $\beta_c=\beta_c(\pi/3)$ and the expected values of the Binder cumulant for cross-over,
first order, second order (3D Ising universality class) and triple point. Bottom: the distribution of the
Polyakov phase as measured on the ensembles generated on the Roberge-Weiss line in order of increasing
temperature from left to right.}
\label{fig:5}
\end{figure}

The next task is to compute the transition temperature along the Roberge-Weiss line $\mu_I/T=\pi/3$.
At high temperatures the Polyakov loop is expected to oscillate between $\arg(P)=0$ and $\arg(P)=-2\pi/3$
sectors and generate a bimodal distribution. As we lower the temperature the two peaks should get
closer and eventually merge into one. This is exactly the situation we observe in the bottom
panel of Fig.~\ref{fig:5} where the probability distribution of $\varphi(U)=\Im(P(U)\times e^{i\pi/3})$ is plotted
for a set of increasing temperatures. To determine the location of the transition point more
precisely we compute the Binder cumulant for this quantity:
\beq
b_4(\varphi) = \frac{\av{\varphi^4}}{\av{\varphi^2}^2} \,.
\eeq
The Binder cumulant value is expected to be $3$, $1$, $1.604$, and $1.5$ for cross-overs, first-order
transitions, second order in the $3$D Ising universality class, and triple point respectively.
In the standard scenario, the expectation is that the Binder cumulant curve as a function of
temperature will assume either the value of $1.604$ or $1.5$ as we intersect with the
(pseudo)critical curve at $\beta_c(\pi/3)$. As we can see from Fig.~\ref{fig:5} this
expectation is close to what we observe. However, if we look closely we find that the
intersection point is not very close to neither of these values. Since the statistical
errors are quite small --they are comparable with the thickness of the line in Fig.~\ref{fig:5}--
this must be a finite volume effect. Larger volume simulations are required to establish
the exact nature of the transition and whether the (pseudo)critical line intersects
the Roberge-Weiss line at the same temperature.

To conclude, we analyzed the phase diagram of $N_f=3$ QCD at a pion
mass of $m_\pi=760\MeV$. Our results are consistent with the
Roberge-Weiss first order phase transition terminating at a point
close to the (pseudo)critical curve. In this study we only used
one lattice volume, $6^3\times4$, and we cannot yet distinguish
between a second order transition or a triple point. We plan to
generate ensembles at larger volume and use finite size scaling 
to answer these questions.

\medskip
\noindent{\bf Acknowledgments:} The computational resources for this project were 
provided in part by the George Washington University IMPACT initiative and in part by the $\chi$QCD collaboration. 
This work is supported in part by the NSF CAREER grant PHY-1151648.

\bibliographystyle{JHEP}
\bibliography{my-references}

\providecommand{\href}[2]{#2}\begingroup\raggedright\begin{thebibliography}{10}

\bibitem{Roberge:1986mm}
A.~Roberge and N.~Weiss, {\it {Gauge theories with imaginary chemical potential
  and the phases of QCD}},  {\em Nucl. Phys.} {\bf B275} (1986) 734.

\bibitem{deForcrand:2002ci}
P.~de~Forcrand and O.~Philipsen, {\it {The QCD phase diagram for small
  densities from imaginary chemical potential}},  {\em Nucl. Phys.} {\bf B642}
  (2002) 290--306, [\href{http://xxx.lanl.gov/abs/hep-lat/0205016}{{\tt
  hep-lat/0205016}}].

\bibitem{DElia:2009qz}
M.~D'Elia and F.~Sanfilippo, {\it {The Order of the Roberge-Weiss endpoint
  (finite size transition) in QCD}},  {\em Phys.Rev.} {\bf D80} (2009) 111501,
  [\href{http://xxx.lanl.gov/abs/0909.0254}{{\tt arXiv:0909.0254}}].

\bibitem{DElia:2009tm}
M.~D'Elia and F.~Sanfilippo, {\it {Thermodynamics of two flavor QCD from
  imaginary chemical potentials}},  {\em Phys.Rev.} {\bf D80} (2009) 014502,
  [\href{http://xxx.lanl.gov/abs/0904.1400}{{\tt arXiv:0904.1400}}].

\bibitem{deForcrand:2010he}
P.~de~Forcrand and O.~Philipsen, {\it {Constraining the QCD phase diagram by
  tricritical lines at imaginary chemical potential}},  {\em Phys.Rev.Lett.}
  {\bf 105} (2010) 152001, [\href{http://xxx.lanl.gov/abs/1004.3144}{{\tt
  arXiv:1004.3144}}].

\bibitem{DElia:2002gd}
M.~D'Elia and M.-P. Lombardo, {\it {Finite density QCD via imaginary chemical
  potential}},  {\em Phys. Rev.} {\bf D67} (2003) 014505,
  [\href{http://xxx.lanl.gov/abs/hep-lat/0209146}{{\tt hep-lat/0209146}}].

\bibitem{DElia:2004at}
M.~D'Elia and M.~P. Lombardo, {\it {QCD thermodynamics from an imaginary
  $\mu_B$: Results on the four flavor lattice model}},  {\em Phys.Rev.} {\bf
  D70} (2004) 074509, [\href{http://xxx.lanl.gov/abs/hep-lat/0406012}{{\tt
  hep-lat/0406012}}].

\bibitem{DElia:2007ke}
M.~D'Elia, F.~Di~Renzo, and M.~P. Lombardo, {\it {The strongly interacting
  quark gluon plasma, and the critical behaviour of QCD at imaginary $\mu$}},
  {\em Phys.Rev.} {\bf D76} (2007) 114509,
  [\href{http://xxx.lanl.gov/abs/0705.3814}{{\tt arXiv:0705.3814}}].

\bibitem{Nagata:2011yf}
K.~Nagata and A.~Nakamura, {\it {Imaginary Chemical Potential Approach for the
  Pseudo-Critical Line in the QCD Phase Diagram with Clover-Improved Wilson
  Fermions}},  {\em Phys.Rev.} {\bf D83} (2011) 114507,
  [\href{http://xxx.lanl.gov/abs/1104.2142}{{\tt arXiv:1104.2142}}].

\bibitem{Ferrenberg:1989ui}
A.~M. Ferrenberg and R.~H. Swendsen, {\it {Optimized Monte Carlo analysis}},
  {\em Phys. Rev. Lett.} {\bf 63} (1989) 1195--1198.

\bibitem{Alexandru:2010yb}
A.~Alexandru and U.~Wenger, {\it {QCD at non-zero density and canonical
  partition functions with Wilson fermions}},  {\em Phys.Rev.} {\bf D83} (2011)
  034502, [\href{http://xxx.lanl.gov/abs/1009.2197}{{\tt arXiv:1009.2197}}].

\bibitem{Nagata:2010xi}
K.~Nagata and A.~Nakamura, {\it {Wilson Fermion Determinant in Lattice QCD}},
  {\em Phys.Rev.} {\bf D82} (2010) 094027,
  [\href{http://xxx.lanl.gov/abs/1009.2149}{{\tt arXiv:1009.2149}}].

\bibitem{Alexandru:2005ix}
A.~Alexandru, M.~Faber, I.~Horv\'ath, and K.-F. Liu, {\it {Lattice QCD at
  finite density via a new canonical approach}},  {\em Phys.Rev.} {\bf D72}
  (2005) 114513, [\href{http://xxx.lanl.gov/abs/hep-lat/0507020}{{\tt
  hep-lat/0507020}}].

\bibitem{Li:2011ee}
A.~Li, A.~Alexandru, and K.-F. Liu, {\it {Critical point of $N_f = 3$ QCD from
  lattice simulations in the canonical ensemble}},  {\em Phys. Rev.} {\bf D84}
  (2011) 071503, [\href{http://xxx.lanl.gov/abs/1103.3045}{{\tt
  arXiv:1103.3045}}].

\end{thebibliography}\endgroup

\end{document}